\documentclass{aa}
\usepackage{graphicx}
\usepackage{setspace}
\usepackage{longtable,lscape}
\usepackage{hhline}
\usepackage{array}
\usepackage{natbib}
\usepackage{footnote}
\usepackage{txfonts}
\title{Evidence of a substellar companion around a very young T Tauri star
\thanks{Based on observations collected with the CRIRES spectrograph at the 
VLT/UT1 8.2-m Antu Telescope (ESO runs ID 385.C-0706(A) and 093.C-0400(A)) at the Paranal Observatory, Chile.}}
\author{P. Viana Almeida
          \inst{1,3}
	 \and
	  J. F. Gameiro
		\inst{2,3}
	  P. P. Petrov
	  \inst{5}
	\and
          C. Melo
	  \inst{4}	
	\and
          N. C. Santos
	  \inst{2,3}
	\and
          P. Figueira
	  \inst{2,3}
	\and
	  S. H. P. Alencar
		\inst{1}
          }
\institute{Universidade Federal de Minas Gerais, 31270-901 Pampulha, Belo Horizonte - MG, Brasil\\
	  \email{palmeida@fisica.ufmg.br}
	  \and
	     Departamento de F\'{\i}sica e Astronomia, Faculdade de Ci\^{e}ncias, Universidade do Porto, Rua Campo Alegre, 4169-007 Porto, Portugal
	  \and
	     Instituto de Astrof\'{\i}sica e Ci\^{e}ncias do Espaço, Universidade do Porto, CAUP, Rua das Estrelas, 4150-762 Porto, Portugal
	  \and
             ESO, Alonso de Cordova 3107, Casilla 19001, Vitacura, Santiago, Chile             
	  \and
	    Crimean Astrophysical Observatory, Russian Academy of Sciences, 298409, Nauchny, Republic of Crimea
		   }
  \abstract
   {We present results from a Near Infrared multi-epoch spectroscopic campaign to detect a young low-mass companion 
to a T Tauri star.
AS 205A is a late-type dwarf ($\approx$K5) of $\sim$ 1 M$\sun$ that belongs to a triple system.
 Independent photometric surveys discovered that AS 205A has two distinct periods
(P$_1$=6.78 and P$_2$=24.78 days) detected in the light curve that persist over several years. 
Period P$_1$ seems to be linked to the axial-rotation of the star and is caused by the presence of cool surface spots. 
Period P$_2$ is correlated with the modulation in AS 205A brightness (V) and red color (V-R), 
consistent with a gravitating object within the accretion disk. 

We here derive precise Near Infrared radial velocities to investigate the origin of period P$_2$ 
which is predicted to correspond to a cool source in a Keplerian orbit with a semi-major axis of $\sim$0.17 AU 
positioned close to the inner disk radius of 0.14 AU. 
The radial velocity variations of AS 205A were found to have a period of P $\approx$ 24.84~days
and a semi-amplitude of 1.529 kms$^{-1}$. 
This result closely resembles the P $_2$ period in past photometric observations (P $\approx$ 24.78~days).
The analysis of the cross-correlation function bisector 
has shown no correlation with the radial velocity modulations,
strongly suggesting that the period is not controlled by stellar rotation. Additional activity 
indicators should however be explored in future surveys.
Taking this into account we found that the presence of a 
substellar companion is the explanation that best fits the results. 
We derived an orbital solution for AS 205A
and found evidence of a m$_{2}$sin$i$ $\simeq$19.25~M$_{Jup}$
 object in an orbit with moderate eccentricity of e $\simeq$~0.34.   
If confirmed with future observations, preferably using a multiwavelength survey approach, 
this companion could provide interesting constraints on brown dwarf and planetary formation models.}

\begin{document}
\maketitle
\section{Introduction}
To best address the planetary formation process we should study pre-main-sequence objects (PMS)
that are still enshrouded in their original environmental conditions. 
The age range between $\sim$1-5~x~10$^6$ yrs in PMSs is particularly interesting since it is the expected moment when the physical conditions 
in the disk enable planetesimal growth and disk migration.  \\
\indent In the last few years, several young planet detection surveys have been conducted in PMS stars using
different techniques, such as direct imaging \citep[e.g.,][]{fenetre,kraus12} 
or radial velocity (RV) monitoring \citep[e.g.,][]{martin06,setii,croqui12}.
To deploy RV surveys on young low-mass stars (which are considerably faint, distant and affected by extinction) 
is, in most of cases, an arduous task.
This is partly due to the presence of cool spots produced by strong stellar magnetic fields \citep[e.g.,][]{johni,melo03}, 
and to intense stellar activity which causes RV variability of PMS objects. 
These effects introduce high uncertainties to 
the RV measurements as well as relevant modulations of spectral line
profiles that can mimic the presence of substellar companions \citep[e.g.,][]{saar97,hatz02,fi10_2}.
While spectral distortions produced by cool spots are wavelength dependent \citep[e.g.,][]{vrba,hue98,fi10_2}, RV variations
caused by the presence of a low-mass companion affect all wavelengths equally. 
To diagnose whether RV variability is companion- or spot-induced, 
 RVs and cross-correlation function bisector (BIS) measurements are correlated 
as are measures of other activity indices \citep[e.g.,][]{figueira2013}
in order to search for activity induced trends (see Sect 4.2). \\
\indent In the Near Infrared (NIR), the effect of stellar spots is expected to be considerably weakened when compared to the 
optical domain \citep[however, see][]{rei13}. Furthermore, extinction of PMS stars 
is also significantly reduced in the NIR. For these reasons, RV studies on low-mass PMS stars 
can be delivered with an enhanced precision and a higher signal-to-noise ratio (S/N) in the NIR.
Observations in the NIR also favor the detection of substellar companions to T Tauri stars (TTS). 
TTS are young solar analogs ($\sim$1-10~x~10$^6$ yrs) that have spectral energy distribution (SED) 
emission that peaks at 1-2 $\mu$m. Classical TTSs (CTTS)s are a subclass that further display observational evidence 
of the presence of an accreting circumstellar disk \citep[e.g.,][]{bou07}. 
It is in the rich circumstellar environment of CTTSs
that protoplanets are thought to coalesce and grow in size \citep[e.g.,][]{poli}.
More recently, this idea has been explicitly reinforced by the findings of \cite{donati2016, johnskrull2016} and \cite{mann2016}, 
who reported robust evidence of the existence of Jupiter- and Neptune-sized objects around young TTS (2 Myrs and 5-10 Myrs, respectively). \\
\indent One of the the most promising CTTS for protoplanetary searches is AS 205A ($\approx$V866 Sco). 
It is a young ($\sim$ 0.5 Myrs) late-type dwarf ($\approx$K5) with mean
V$=$12.4 mag 
that belongs to a hierarchical triple system. 
At an angular separation of 1.3$\arcsec$ ($\approx$180 AU at 140 pc) from
AS 205A lies a very low-mass (K7/M0) spectroscopic binary \citep[][]{ghez93, pra03, eisner05}. 
An extensive photometric survey by \cite{art10} revealed two distinct and stable periods
(P$_1$=6.78 and P$_2$=24.78 days) in the power spectrum of the light variations of AS 205A. 
The 24-day photometric period was also confirmed by \cite{percy10}. 
The value of P$_1$ is typical for a rotational period of
a CTTS and is caused by the presence of cool surface spots. 
In this context, the anti-phase variations of the (\textit{U-B}) color can be explained as
chromospheric emission related to the cool spots. The phase diagram for P$_2$, 
on the other hand, shows modulation in brightness and red colors, which indicates the presence of a cool source. 
Since AS 205A is about 2 mag brighter than AS 205B \citep[][]{her88} in the V band 
it was concluded that the observed modulated signal (V~=~0.25 mag) belongs to the primary or to its circumstellar environment.
The mass of AS 205A, as derived from its stellar temperature and bolometric luminosity \citep{andrews09}, and
using the recently released pre-evolutionary tracks from \cite{baraffe2015}, is expected to be close to solar ($\sim$ 0.9 M$\sun$). 
According to \cite{art10},
the period P$_2$ should correspond to the Keplerian location of an unknown close companion of AS 205A
which perturbed the accretion disk with density waves.
The orbit was predicted to have semimajor axis of $\sim$0.18 AU, 
which is close to the inner disk radius R$_{in}$=0.14 AU, measured by IR-interferometry
\citep[][]{eisner05, andrews09}. \cite{art10} further interpret the light variations of P$_2$ as the effect of 
scattering or extinction in the disturbed disk near the dust sublimation radius. \\
\indent In this work we derived precise multi-epoch NIR RVs to investigate the origin of period 
P$_2$ and searched for the presence of a low-mass companion. 
We report here our results and briefly discuss the implications of the study.
\section{Observational method and data reduction}
\indent High-resolution observations of AS 205A were conducted in the NIR, most specifically in the H-band where CTTS
photospheric information can be accessed.
Spectroscopic observations were carried out
in good seeing conditions ($\sim$0.8'') between Apr 22, 2010, and May 1, 2014, 
using CRIRES, the NIR high-resolution
spectrograph mounted on the UT1 telescope at Paranal Observatory (ESO). 
AS 205A was visited eight times during this period.
Spectra were collected using the 4096 $\times$ 512 pixel Aladdin III detectors and a 
0.2'' slit which delivered a R$\sim$100000 
around 1598.0 nm (CRIRES setup 36). This particular setting 
was chosen because we could benefit from the  
 CO$_2$ telluric atmospheric lines as simultaneous 
wavelength calibrators \citep[][]{hue08, fig10} to derive precise RVs.   
Previous studies have indeed shown that telluric lines are steady RV zero-point tracers \citep[][]{figueira2010_3} 
that display a long-term stability and, for this reason, can deliver RVs with a precision down to 5-10 ms$^{-1}$. 
Final data were acquired in 2 AB nodding cycles with an average S/N $\sim$20 per pixel and an expected 
final precision of around 30 ms$^{-1}$,
sufficient, according to our simulations, 
to spot an object of a few to several Jupiter masses  
at the estimated photometric period of 24.78 days. \\
\indent In addition to the AS 205A data, we also collected spectra of telluric standard stars (featureless early-type B stars)
obtained with similar airmass and instrumental setup to AS 205A. 
All observations were taken back-to-back with a RV standard HD192310 (K2V) of similar spectral type
to examine the precision of our RV measurements along the time span of our survey.\\
\indent Reduction was performed using an optimized IRAF-based pipeline \citep[see][for more details]{vi12}. 
In summary, all spectra were nonlinearity-corrected, dark-current subtracted, flat-fielded, sky-subtracted 
(through subtraction of opposing nodded spectra) and optimally extracted using the \cite{h086} algorithm. 
Owing to the blending of telluric and stellar lines in the final spectra, we also  
performed telluric removal by dividing each extracted spectrum by the spectrum of telluric standards stars. \\
\indent Wavelength calibration was performed using telluric absorption lines 
\citep[see e.g.,][]{fig10, bai12}. 
Laboratory wavelength zero-points of each telluric spectral line were collected 
from the HITRAN database \citep[][]{hitran}.
The RVs in this study were derived using the \cite{fig10} pipeline and its adapted version to young stars
developed by \cite{vi12}.
Both versions of the pipeline are based on a two-dimensional (2D) cross-correlation function (CCF)
 inspired by TODCOR \citep{maz92}. They were specifically built to derive the RV of an object 
relative to the zero-point established by the telluric lines. \\
\indent To determine the barycentric RVs we cross-correlated 
each nodded spectrum against 
the spectra supplied by the NIR PHOENIX synthetic database \citep{hus13} and 
against the spectra of HD192310 observed on the same date. 
Two-dimensional CCFs were fitted using Gaussian function profiles. 
Wavelength solutions, barycentric Julian dates and RVs were then adjusted to the center of mass of the solar system using
the \cite{bret88} ephemerids. 
\vspace*{-0.4cm}
\section{Analysis} 
\subsection{Rotation and stellar models}
\indent To obtain the best-fitting template for cross-correlation and determine the projected equatorial velocity
 ($v$~sin~$i$) of the target
we used the effective temperature/ SpT scales of \cite{lulu} to select a subset of PHOENIX models
 compatible with the photospheric properties of AS 205A. 
From the PHOENIX database we collected models with effective temperatures (T$_{eff}$)
ranging from 4100 to 4700~K, surface gravities 
(log $g$) between 3.5 and 4.0 (typical values found in TTS), and 
 metallicities ([Fe/H]) ranging from -0.5 to 0.5 dex. 
The abundances of $\alpha$ elements were considered solar.\\
\begin{figure}[!ht]
\centering
\includegraphics[scale=0.71]{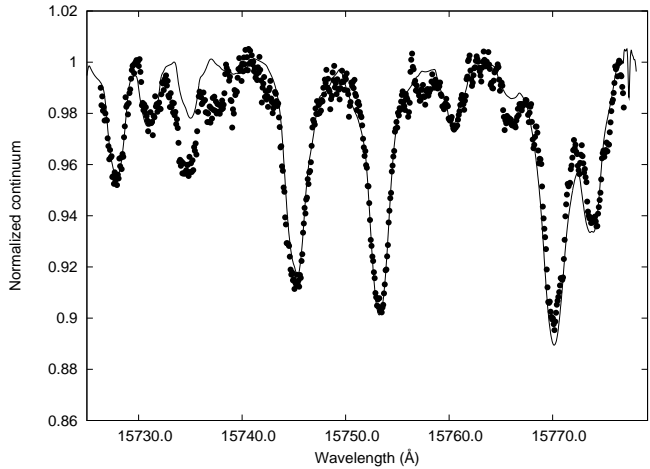}
\caption{Overplot of a stellar synthetic spectra (solid line) convolved with a $v$~sin~$i$ of 11~kms$^{-1}$ 
on a spectrum of AS 205A (dots). 
Data shown were smoothed for clarity, continuum normalized 
and Doppler corrected to the rest wavelength.}
\label{espectro}
\end{figure}
\indent We then used the \textit{gauss} task from the IRAF package to 
degrade the resolution 
of the synthetic models (R$\sim$500000) in order to match 
that of our observations. 
Using a stellar rotation broadening kernel, we applied $v$~sin~$i$, which 
ranged from 6 to 20 kms$^{-1}$, to the models 
creating a wider grid of models. 
We finally used a routine to interpolate through the models to find the model that
minimized the chi-square ($\chi^2$) statistics and
 best adjusted the spectrum of AS 205A. 
Figure~\ref{espectro} shows one of the best-fit models. The final stellar parameters 
of our best model were T$_{eff}$~=~4300~K,
 log $g$~=~3.5~cms$^{-2}$, [Fe/H]~=~0.0 dex, and $v$~sin~$i$~=~11~kms$^{-1}$. \\
\indent Keeping these parameters in mind, if we assume a TTS radius of $3.7R_{\sun}$ \citep[as provided in][]{and10} and a system 
inclination of $\sim$23.6$^{\circ}$ \citep[given by][]{art12} we calculate a rotational 
period of P$\sim$ 6.81 days, which is in close agreement with
previous results for P$_1$ from \cite{art12}. Curiously, the 
$v$~sin~$i$ found by iteratively fitting the models  
 is also close to the estimates used in Artemenko's publication.
It can be seen that the rotational period is clearly not of the order of the
period P$_2$ of 24.78 days encountered in the studies of \cite{art10} and \cite{percy10}. 
The apparent disparity between the two periods may seem to reinforce the idea that P$_2$ 
is of companion origin and not rotationally driven,
 but a possible relation between P$_1$ and 4xP$_2$ cannot be completely ruled out and should be investigated further in the future
with a larger data set.
\subsection{Radial velocities}
\begin{figure}[t]
\centering
\vspace*{-.3cm}
\includegraphics[scale=0.45]{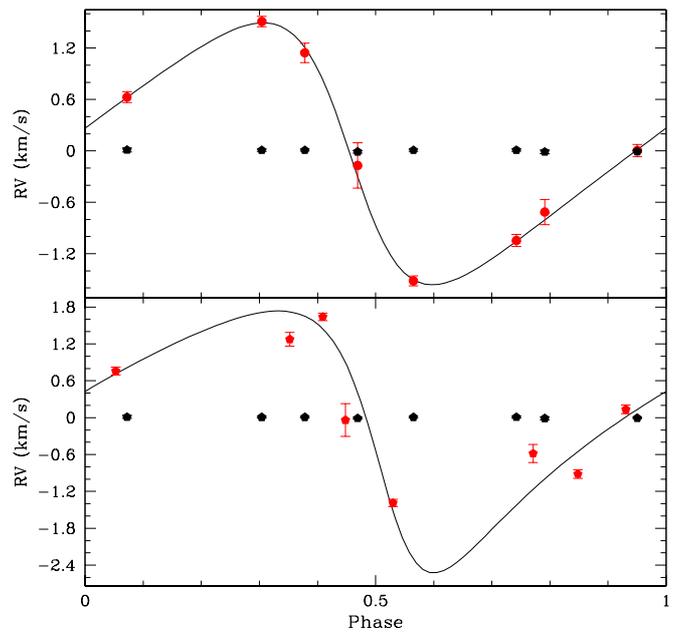}
\caption{Upper panel: RV measurements of AS 205A vs. orbital phase in Julian Days with an orbital period of 24.84 days. 
 Lower panel:
Same but using an orbital period of 24.78 days. Overlayed are the best-fitting Keplerian curves for each period. 
Black crosses close to the x-axis represent RV data from HD192310.}
\label{allspec3}
\end{figure}
The RV measurements for each date are shown in Table~\ref{kstars} and plotted 
in Fig.~\ref{allspec3}. They are the result of the correlation 
of the spectral information
 sampled in Detectors 1, 2, and 4 of the selected CRIRES setup (containing $\sim$ 15 absorption lines). 
Final RV uncertainties were calculated dividing the rms of each RV measurement
 by the square root of the number of independent exposures acquired in each date. The average error bar
of the data is $\sim$85ms$^{-1}$. The reason for this large value is essentially due to the 
increased level of blending of stellar and telluric lines in some of the spectra. 
Telluric removal in these cases is less efficient and is known to decrease 
the S/N of the final spectra, hence the final precision of the measurements.\\
%
%
\setlength\extrarowheight{2pt}
\begin{table}[htbp]
\centering   
\vspace*{0.5cm}                  
\begin{tabular}{cccc}        
\hline\hline                 
MJD &v\textit{rad} [km s$^{-1}$]  & \textless$\sigma$\textgreater \\
\hline
%
2455308.8595 & -1.518 & 0.059  \\
2455403.6020 & 1.144 & 0.073 \\
2455455.5253 & -0.170 & 0.212 \\
2455463.5396 & -0.714 & 0.066  \\
 2455467.4938 & 0.004 & 0.070 \\
2455470.5179 & 0.628 & 0.064 \\
2456767.9046 & 1.510 & 0.062 \\
2456778.7772 & -1.046 & 0.070 \\
\hline 
\end{tabular}
\vspace*{0.5cm}
\caption{Barycentric julian dates of each nodded image, derived radial velocities and final dispersion for each date AS 205A was observed.}
\label{kstars}      
\end{table}

\indent As mentioned before, in order to check for 
possible drifts in the long-term RV measurements obtained using our methodology, we
 retrieved RVs of a control star (HD192310) over the 
time span of the study. In order to keep the same 
intrumental profile and reproduce the same observation conditions as 
in our scientific target,
 spectra of HD192310 were observed back-to-back with AS 205A.
\cite{pepi11}, using high-resolution spectroscopy, 
published the Keplerian solutions of two orbiting planets around HD192310.
The dispersion of their RV measurements, however, had a rms of only 2.6~ms$^{-1}$ 
over a 6.5-year interval. 
Our own study detected variations in the HD192310 
RV profile with a standard deviation of $\leq$11~ms$^{-1}$ over
4 years of observations.
Although it was beyond our detection limit we tried 
nevertheless to adjust the Keplerian solutions of \cite{pepi11}
to the variations observed. The lack of consistent fits implied that 
the dispersion in our RV data was instead associated with the internal 
uncertainty of our method. \\
\indent For the sake of clarity, 
we plot the RV measurements of HD192310 along with those of AS 205A 
in Fig.~\ref{allspec3}.
As we can see in this figure, the RV variations 
of HD192310 over the entire study are comparatively 
small, providing solid evidence 
that observational uncertainties are not at the origin of the RV 
variations observed in the classical T Tauri stars. 
The RVs of HD192310 seem therefore to confirm the stability and precision of the results obtained for AS 205A. \\ 
\vspace*{-0.5cm}
\section{Discussion}
\subsection{Preliminary orbital parameters}
\indent Using the RVs from Table~\ref{kstars} 
we found preliminary Keplerian orbital solutions for the suggested companion. 
For these calculations we considered  
a primary mass as estimated from the recently released pre-evolutionary tracks of
 \cite{baraffe2015} which present significant improvements over older tracks 
commonly employed in previous studies on AS 205A, 
such as those from \cite{siess2000}.
We used the Systemic code\footnote[2]{www.oklo.org} to fit a Keplerian solution
to the RV measurements. To find the best orbital parameters, first we fixed the Period at 24.78 days 
and the eccentricity \textit{e} at null value while letting all the other parameters converge. \\
%
\setlength\extrarowheight{3pt}\vspace*{0.5pt}
\begin{table}[!ht]
\centering                     
\begin{tabular}{l c}        
\hline                
\hline
%
Orbital parameters& Value\\
\hline
P$_{orb}$ (days)  & 24.84 $\pm$ 0.03 \\
m$_{1}$ (M$_\sun$)         & \textbf{0.9}\\
m$_{2}$sin$i$ (M$_{Jup}$) & \textbf{19.25} $\pm$ 1.96 \\
$K$ (kms$^{-1}$) & 1.529 $\pm$ 0.16 \\
$e$ & 0.34 $\pm$ 0.06 \\
$\omega$ (deg) & 94.14$\pm$ 7.67 \\
Semi-major axis (AU) & 0.162 $\pm$ 0.04 \\
V$_{sys}$ & -10.25 $\pm$ 0.07 \\
\hline
$\chi^2_{red}$ & 1.07 \\
\textit{O - C} (ms$^{-1}$) & 73.4\\
\hline 
\end{tabular}
\vspace*{0.3cm}
\caption{Orbital parameters for the substellar companion.}
\label{orbital}
\end{table}

\indent As soon as we obtained preliminary approximations for the systemic 
velocity V$_{sys}$, longitude of periastron $\omega$ and m$_{2}$sin$i$, 
we fine-tuned the results by letting all the orbital 
parameters converge simultaneously. In Fig.~\ref{allspec3} we present 
the best Keplerian curves for photometric
period P$_2$ from \cite{art10} and the period that best-fitted our RV data.
The overall rms and the reduced chi-quare ($\chi^2_{red}$) of the solution using a 
period of 24.78~days were of 421~ms$^{-1}$ and 66.1, respectively. Conversely, 
when minimizing the rms of the fit, after some iterations, we obtain a period of 24.84 days with a 
much lower rms of 73.4~ms$^{-1}$ and a
 $\chi^2_{red}$ of 1.07. 
This $\chi^2_{red}$ of the P$\sim$24.84~day solution found for the RV data  
suggests a good fit to the data. Therefore, it seems that this period better explains the results.
In Table~\ref{orbital} we display the orbital elements for the 24.84 day period final fit.
The uncertainties depicted are confidence intervals provided by the bootstrap method. 
We resampled the original data set 100000 times and fitted a Keplerian solution 
to each resampled set thereby 
producing distributions for each orbital parameter. 
 We note that 
the degrees of freedom of the orbital solution almost equals the number of data points. 
This implies that we must be cautious when interpreting of the results.  
\subsection{Stellar activity}

\begin{figure}[t]
\centering
\includegraphics[scale=0.42]{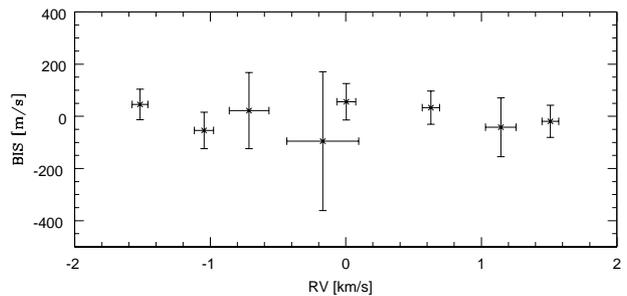}\vspace*{-4cm}
\caption{BIS plotted against RVs of AS 205A. As we can see no correlation was detected.}
\label{bis}
\end{figure}

\indent To determine whether variability was driven by surface cool spots, 
we probed the correlation of the BIS and the RVs in the CCFs obtained in each date. 
A clear correlation between these two quantities 
is expected when CCF profiles are subject to asymmetries caused by stellar spots \citep[][]{quelo01}. 
Since young active stars are likely to have BIS that vary with time, their study can provide useful
 hints on the activity of AS 205A at the time of the observations. Furthermore, high
 stellar $v$~sin~$i$ (11kms$^{-1}$ in the AS 205A case) are expected 
to favor the sensitiveness of the BIS 
method as an activity-induced indicator \citep[see e.g.,][]{san03}. 
We note, however, that 
in the wavelength interval of our observations
this correlation is expected to decrease by a factor of $\sim$~2-3 \citep[see, e.g., models from][]{deso07,rei10, mage12}. 
We show in Fig.~\ref{bis} the BIS plotted against the RVs obtained for AS 205A.
Despite the small sample size we performed a preliminary 
examination of the correlation between RV and BIS in our data set using different statistical diagnostics. 
We used a Bayesian analysis tool
 from \cite{figueira2016} to compute Pearson's correlation coefficient $\varepsilon$ and Spearman's rank correlation coefficient 
$\rho$ and associated
p-values (for more details on the method see the above publication). The Pearson's coefficient, which measures 
the strength of the linear association between the two variables, was found to be
-0.171 with a 0.686 two-sided p-value and the Spearman's rank coefficient, 
which characterizes their monotonic relationship, was found to be
approximately -0.071 with a 0.867 two-sided p-value.
Both results seem to suggest that there is no statistical correlation between 
the distributions of the two quantities.
 These results led us to conclude 
that, at least in the NIR, the derived RVs are not correlated with BIS, hence, most probably they 
are not driven by stellar activity. \\
\indent In the optical domain, however, past RV measurements may imply the influence of 
cool spots in the shape of the CCFs. \cite{melo03} derived six  
 optical RVs (\textless$\sigma$\textgreater$\sim$450ms$^{-1}$)
for AS 205A using the CORALIE spectograph and reported 
clear RV variations on the order of 4.34 kms$^{-1}$. Strikingly, this variability is 
comparable in order of magnitude to our own RV variations.
We note, however, that one of its AS 205A CCFs \citep[shown in Fig. 5 of][]{melo03}
displayed an abnormal deformation which the author interpreted as a 
possible spectroscopic stellar companion. We suspect that this CCF deformation
may have been caused, in part, by stellar activity even 
though visual inspection of its CCF seems to rule out this explanation.\\
\indent Future studies should consider further testing of 
the effects of stellar activity and, more specifically, of rotational harmonics in the RV measurements.  
Harmonics are expected to affect RVs when 
rotational modulation of light by stellar spots occurs \citep[see, e.g.,][]{boisse11}.
 We made a preliminary test of the presence of such rotational harmonics in our data 
(using the rotational period 
calculated in Sect. 3.1.) by fitting sinusoidal functions to the residuals
 of the Keplerian fits shown in Fig~\ref{allspec3}.  
We were unable, however, to find any coherent signs 
of the first three rotational harmonics 
(P$_{\text{rot}}$/2, P$_{\text{rot}}$/3, P$_{\text{rot}}$/4) which could be imprinted in the
RV jittering amplitudes. This result and the goodness of the fit of the adjusted Keplerian solution 
using P~=~24.84 days 
increases our confidence that the results presented do not depend on stellar spot modulation. 
But to further explore these effects
we need a better sampling of the RV curve with a higher number 
of data points. This kind of study would certainly improve 
the orbital charaterization of the reported substellar companion.
\\ 
\subsection{New perspective on the AS 205A system}
\indent Following the results presented here and the results from \cite{art10}, we 
suggest the existence of a low-mass companion to AS 205A.
We propose that observational evidence of this low-mass companion in previous campaigns 
 could have been hindered by the prolific IR emission observed in AS 205A. \\
\indent Assuming that the orbit of the planet is aligned with the stellar rotation
and that the inclination of the system is between 
 $\sim$25~$^{\circ}$ \citep{and10, art12, ponto11} and 
$\sim$15~$^{\circ}$, as the study of 
\cite[][]{salyk14} suggests, then the absolute mass range 
of the proposed substellar companion would be between $\sim$45.6 and 74.3 M$_{\text{Jup}}$.
This mass range suggests an upper and lower limit in T$_{eff}$ of around $\sim$2200 K and 
1500K \citep[see, e.g.,][]{kirk05,rice2010}, respectively. 
In this temperature range the object would fall into the 
 temperature-mass domain of ultracool dwarfs or very late-type M stars.
Since the peak emission of such objects is located
in the NIR range, close to 1.0 $\mu$m \citep[][]{cush06,sarro2013}, its SED
 could thus be masked by the strong 
circumstellar IR emission of AS 205A \citep[see, e.g.,][]{and10}. 
Ideally, new dedicated spectroscopic observations, both high-resolution and high S/N, 
should be prepared in the the near future with the aim of identifying notable spectral features 
of brown dwarfs or very late-type M stars that could be imprinted 
in the AS 205A spectrum. Given the very distinct nature of late-type M stars and brown dwarfs 
\citep[see, e.g.,][]{maclean2007, rice2010}, this strategy 
could lead to a successful identification of the spectral type and mass regime of the substellar companion.\\
\indent Recent sensitive high-resolution mm observations of AS 205A 
with the ALMA facility \citep[][]{salyk14} revealed an intriguing extended asymmetric profile of the $^{12}$CO (2-1) emission. 
Despite their best efforts, 
the authors could not explain this emission
merely on the basis of a Keplerian gas-disk models and/ or stellar wind parametrization. Indeed, when compared
 with typical outflows produced in protostars \citep[e.g.,][]{jorgensen}, the $^{12}$CO emission of 
AS 205A presents quite distinct and unique characteristics \citep[see, e.g.,][]{salyk14}. 
While some of the emission can be associated with tidal stripping from 
the close binary AS 205B at just 1.3~$\arcsec$, it is unlikely that this phenomenon alone could explain 
the observations. 
Arguably, the discrepancies observed may suggest the unaccounted presence of an additional companion to AS 205A. 
This possibility should be explored in the future if follow-up 
studies confirm the existence of a close substellar object. 
\section{Conclusion}
\indent It is interesting to note the close agreement between the estimated
 photometric period P$_2$ 
($\sim$24.78 days) of \cite{art10} 
and the Keplerian period obtained with our NIR spectroscopy 
(P$\sim$24.84~days). Period P$_2$, to the best of our knowledge,
 does not seem to be 
a rotational modulation produced by spots in the stellar surface. 
More RV measurements, however, are necessary in order to confirm these results. 
Even though unacounted phenomena such as stellar activity and/ or accretion \citep[\textit{e.g.,}][]{bou07} could have
 influenced the spectral profile,
the high-precision attained in this study (below 70~ms$^{-1}$ in most of the cases) and 
the lack of correlation of BIS and RVs variations provide strong support for our companion interpretation. 
The substellar companion hypothesis is, to the best of our knowledge, 
the most robust explanation for the results obtained.\\
\indent The confirmation of a substellar companion at the border of the inner gap of the protoplanetary disk 
at such an early time ($\sim$0.5~-~1 Myrs)
could provide interesting constraints on brown dwarf and planet formation theories and on 
the process in which planets migrate within the circumstellar disk. 
The AS 205A system remains an important target to engage new observations in the near future.  \\
\begin{acknowledgements}
PVA acknowledges the support from Comiss\~{a}o Nacional de Pesquisa (CNPq) and the 
Brazilian National Council of Scientific and Technologic Development,
in the form of a grant PDJ with reference 160111/2012-9. 
This work was supported by Funda\c{c}\~ao para a Ci\^encia e a Tecnologia (FCT) 
within projects reference PTDC/FIS-AST/1526/2014 (POCI-01-0145-FEDER-016886) 
and UID/FIS/04434/2013 (POCI-01-0145-FEDER-007672).
PF and NCS acknowledge support by Funda\c{c}\~ao para a Ci\^encia e a Tecnologia (FCT) through Investigador 
FCT contracts of reference IF/01037/2013 and IF/00169/2012, respectively, and POPH/FSE (EC) by FEDER funding 
through the program ``Programa Operacional de Factores de Competitividade - COMPETE''. 
PF also acknowledges support from Funda\c{c}\~ao para a Ci\^encia e a Tecnologia (FCT) 
in the form of an exploratory project of reference IF/01037/2013CP1191/CT0001.
\end{acknowledgements}
\bibliography{maria}
\bibliographystyle{aa}
\end{document}